\documentclass[11pt]{article}
\usepackage{moriond,epsfig}

\def\IJMP{{\em Int. J. Mod. Phys.}  A }
\def\MPL{{\em Mod. Phys. Lett.}  A }

\def\NPA{{\em Nucl. Phys.} A }
\def\NPB{{\em Nucl. Phys.} B }
\def\MPL{{\em Mod. Phys. Lett.}  A }
\def\PLB{{\em Phys. Lett.}  B }
\def\PRL{{\em Phys. Rev. Lett. }}
\def\PRD{{\em Phys. Rev.} D }

\def\be{\begin{equation}}
\def\ee{\end{equation}}
\def\bea{\begin{eqnarray}}
\def\eea{\end{eqnarray}}

\begin{document}
\begin{flushright}
TIFR/TH/00-57 \\
hep-ph/0010048 \\
\end{flushright}
\vspace*{3cm}
\title{HEAVY ION PHYSICS : THEORY}

\author{R.\ V.\ GAVAI}

\address{Department of Theoretical Physics, Tata Institute of
Fundamental Research, \\ Mumbai 400 005, India} 

\maketitle\abstracts{Lattice quantum chromodynamics (QCD) predicts
a new state of matter, called quark-gluon plasma (QGP), at
sufficiently high temperatures or equivalently large energy densities.
Relativistic heavy ion collisions are expected to produce such energy
densities, thereby providing us a chance to test the above prediction.
After a brief introduction of the necessary theoretical concepts, I 
present some critical comments on the experimental results with an
aim to examine whether QGP has already been observed experimentally.}

\section{Introduction}

The standard model of particle physics has been tested with great
precision at LEP. Most of these tests exploit the fact that the
corresponding coupling is weak and hence usual weak coupling
perturbation theory can be employed in deriving the required theoretical
predictions for them.  The electromagnetic and weak couplings are indeed
rather small in the currently accessible energy range: $\alpha_{em}
\simeq 7.3 \times 10^{-3}$ and $\alpha_w \simeq 3.4 \times 10^{-2}$.
However, the strong interaction coupling, $\alpha_s$, is (i) a strongly
varying function of energy in the same range, (ii) about 0.11 at the
highest energy at which it has been measured so far, and (iii) $\sim 1$
at typical hadronic scales.  Therefore, testing the strongly interacting
sector of the standard model using only perturbation theory is a major
shortcoming of the current precision tests of the standard model. 

Formulating quantum chromodynamics (QCD), which is an $SU(3)$ gauge
theory of quarks and gluons, on a discrete (Euclidean) space-time
lattice, as proposed by Wilson \cite{wil}, and simulating it
numerically, as first done by Creutz \cite{cre}, one can obtain 
several post-dictions \cite{lat} of QCD in the non-perturbative domain 
of large $\alpha_s$. These include both the qualitative aspects like
confinement and spontaneous chiral symmetry breaking, and quantitative details
such as hadron masses and their decay constants. While the latter agree
with the known experimental results within the sizeable theoretical
errors, it is fair to say that no serious experimental test of any
non-perturbative prediction of QCD has so far been made, with the possible 
exception of the D-meson decay constant. Relativistic
heavy ion collisions offer a great window of opportunity to do
so. Application of lattice techniques to finite temperature QCD has
resulted \cite{ftqcd} in the prediction of a new state of matter,
called Quark-Gluon Plasma(QGP), at sufficiently high temperatures or energy 
densities. Chiral symmetry, broken spontaneously at zero temperature,
seems to be restored in this new phase characterised by a much larger
degrees of freedom characteristic of almost ``free'' quarks and
gluons. 

Let me sketch in very brief the reason this prediction needs to be
regarded as a crucial test of QCD in the non-perturbative domain.
Starting from the text-book expression for the partition function,
\begin{equation} 
\cal{Z} = \rm{Tr}~\exp \bigg[ - (\hat H - \mu \hat N) / T \bigg] \ , 
\label{eq:zed} 
\end{equation} 
where $\hat H$ is the Hamiltonian for QCD, $\hat N$ is the baryon number
operator, and the trace is taken over all physical states, various
thermodynamic quantities of interest, such as the energy density or the
pressure at a given temperature $T$ and baryonic chemical potential
$\mu$, can be obtained as expectation values of appropriate operators
with respect to {\cal Z}.  These are computed by first rewriting the
partition function exactly in terms of an Euclidean path integral over
the underlying quark and gluon fields: 
\begin{equation} 
{\cal Z} = \int {\cal D} A_\mu {\cal D} \bar \psi {\cal D} \psi \exp \bigg[ -
\int_0^{1/T} dt \int d^3 x~{\cal L}_{QCD} (A_\mu, \bar \psi, \psi; \mu, g ) 
\bigg] \ . 
\label{eq:zed1} 
\end{equation} 
This expression resembles the corresponding one for QCD at $T$ = 0 a
lot.  It only differs in having a finite extent ($= 1/T$) for 
the (compact) Euclidean time.  This suggests that lattice
techniques can be useful in extracting information on QCD thermodynamics
non-perturbatively. Discretizing the space-`time', and using a gauge
invariant formulation of QCD on lattice, thermal expectation values
can be computed at a finite lattice spacing (or equivalently at a finite
ultra-violet cut-off) using, e.g, numerical Monte Carlo techniques. 
Repeating these calculations for a decreasing sequence of the 
lattice spacings at a fixed physical scale, one can obtain results 
in the desired continuum limit, although it needs massive computational
efforts.

Simulations of lattice QCD at finite temperature thus results in
parameter free information on QCD thermodynamics starting from first
principles, since the only parameters, quark masses and the scale of
QCD, can be fixed from zero temperature simulations.  There are,
however, some caveats and subtleties which make these computations
difficult and nontrivial.  E.g., defining a chiral symmetry on the
lattice with a given number of massless (or light) flavours is still a
thorny subject.  How small a lattice spacing is adequate is not clear.
Nevertheless, the existence of a new chirally symmetric phase seems
\cite{lat,ftqcd} well established.  Furthermore, this phase appears to
be inherently non-perturbative in the experimentally interesting range
of $1 \leq T/T_c \leq 4$-10, where $T_c \sim 170$ MeV is the transition
temperature at which the energy density varies most rapidly. The energy
density, $\epsilon$, in this range is 15-20\% smaller \cite{ber} than
the value of the corresponding ideal gas of quarks and gluons whereas a
maximum of 3-5\% deviation is allowed for a weakly interacting
perturbative QGP. While the precise values for $\epsilon$, or $T_c$, as
well as the nature of the phase transition (whether first order or
second) depend on the number of light quark flavours, the quoted values
above being for 2 flavours of mass about 15 MeV, many simulations with
varying numbers of light flavours suggest that an energy density greater
than 1 GeV/fm$^3$ is needed to reach the QGP phase. 

Collisions of heavy ions at very high energies can potentially produce
regions with such large energy densities. Furthermore, since the
transverse size of such regions is given by the diameter of the
colliding nuclei, one can hope that these collisions will satisfy the
necessary thermodynamical criteria of large volume ($L \sim 2R_A \gg
\Lambda^{-1}_{QCD}$) and many produced particles. A crucial, and as yet
unanswered, question is whether thermal equilibrium will be reached in
these collisions, and if yes, when and how.  A reliable estimate of the
the energy density attained is consequently hard to get.  Assuming 
$i)$ a ``plateau" in the rapidity distribution (in the central region of 
the cm frame) and $ii)$ a ``leading baryon" effect or a baryon-free central 
region, Bjorken \cite{bjo} argued that for sufficiently high energies, 
the colliding nuclei with mass number $A$ bore through each other, leaving
behind a baryonless blob of produced particles in the center (around 
$y_{cm} = {1\over 2} \ln ~[(E + P_L) / (E - P_L)] \sim 0$).  The energy 
density in the blob after an equilibration time $\tau_0$ was estimated by 
him to be 
\begin{equation}
\epsilon = {1\over {\cal A} \tau_0}\  \cdot \ {dE_T \over dy} \ ,
\label{eq:one}
\end{equation}
where the effective area ${\cal A} = \pi R^2_A = 3.94~A^{2/3}$ fm$^2$
and $dE_T/dy$ is the measured transverse energy per unit rapidity round
$y_{cm} \approx 0.0$.   Depending on the value of this initial energy
density and the equation of state, the blob goes through various stages
of evolution such as QGP, mixed phase and hadron gas, as it cools by
expanding. A further rapid expansion of the hadron gas leads to such
large mean free paths for the hadrons that they essentially decouple
from each other.  If this freeze-out is sufficiently fast, the
free-streaming hadrons, $\pi, K, \cdots$ etc. will retain the
memory of the thermal state from which they were born by having thermal
momentum distributions. Thus the information from observables related to
light hadrons can tell us about the temperature at this `thermal
freeze-out' and the velocity of expansion. To get a glimpse at still
earlier times, one has to turn to harder probes which typically
involve larger scales such as masses of heavy quarkonia, as we will
discuss below. 

Although both the Bjorken scenario and Eq.  \ref{eq:one} are 
widely used in all current data analyses seeking to extract information on 
whether QGP did form in those collisions, even the present highest CERN 
collision energies may not be sufficient for either to hold. 
This is mainly because many baryons 
appear to get deposited in the central region of $y_{cm} \approx 0.0$ and the
rapidity distribution also seems to be a Gaussian.  A more reliable analogue of
Eq. \ref{eq:one} is however not available in such a case.  Note that even
the theoretical estimate from lattice QCD above was for a baryonless
case which too may be inapplicable for the present day collisions.  In 
addition to temperature, one needs to increase the baryon density of the
strongly interacting matter or equivalently increase the baryonic
chemical potential $\mu$ to obtain a baryon-rich plasma.  In
principle, one knows how to handle the case of a nonzero baryon density
on the lattice but it has so far turned out to be difficult in practice.
Usual lattice techniques fail for nonzero $\mu$ due to technical
reasons \cite{ftqcd} and attempts to overcome \cite{lat} these have not
been successful either.  Thus a greater theoretical effort is required
to obtain a QCD prediction for the energy density for nonzero
$\mu$, which may be more relevant to the existing heavy ion data, and also
to obtain the analogue of Eq.  \ref{eq:one} in that case.    Of course,
one can in stead go for higher energies to test QCD, where one expects
to obtain a baryon-free region, making both the lattice estimate and Eq.
\ref{eq:one} more accurate descriptions. While this will be done in the
near future at RHIC, BNL and LHC, CERN, the existing data already show
many new and exciting features.

\section{Results from CERN}

The experimental programs of high energy heavy ion collisions are
being pursued actively at present in the Brookhaven National Laboratory
(BNL), New York and CERN, the European Laboratory for Particle
Physics, Geneva. $Au$-$Au$ collisions at $\sqrt{s} = 4.7A$ GeV
$\simeq 0.92$ TeV have been studied at BNL while $Pb$-$Pb$ collisions
at $\sqrt{s} = 17.3A$ GeV $\simeq 3.6$ TeV have been investigated at
SPS, CERN using beams of gold ions at 2.1 TeV/c and lead ions at 32.9 TeV/c 
respectively. Earlier sulphur beam at 6.4 TeV/c was used on sulphur
and uranium targets at SPS, CERN and those results form a benchmark over
which several aspects of $Pb$-$Pb$ collisions have been compared. I will
focus largely on the latter since they correspond to the highest
$\sqrt{s}$ used so far. Due to space restrictions, I will also have to 
restrict myself to highlights and I have to refer the reader for more
details to the proceedings of Quark Matter conferences \cite{qm}. 

\subsection{Initial Energy Density}

The NA49 experiment reported measurements on $dE_T/d\eta$ quite a
while ago \cite{na49} and reported $dE_T/dy \simeq 405$ GeV for $Pb$-$Pb$. 
Using a canonical guess of 1 fm for the formation time, one obtains from
Eq. \ref{eq:one}
\begin{equation}
\epsilon^{Pb-Pb}_{Bj} (1 {\rm fm}) = 2.94 \pm 0.3 {\rm GeV/fm}^3 \ ,
\label{eq:two}
\end{equation}
which is certainly above the characteristic QGP-phase values from
lattice QCD mentioned in Sec. 1. Since appreciable numbers of baryons at 
$y_{cm} \sim 0$ have been observed at SPS, it is doubtful that the current 
energies are high enough for creating a baryon-free region assumed 
for Eq. \ref{eq:one}.  One has to be cautious therefore and make sure that 
other independent estimates are also similar and they do appear to be so.  
Using the lattice results for baryonless plasma of 2 light flavours, the 
above estimate suggests a plasma temperature of about 220 MeV or 
about 1.3 $T_c$.

\subsection{Hadron Yields}
\begin{figure}[htbp]
\epsfxsize=8cm
\epsfysize=8cm
\centerline{\epsfbox{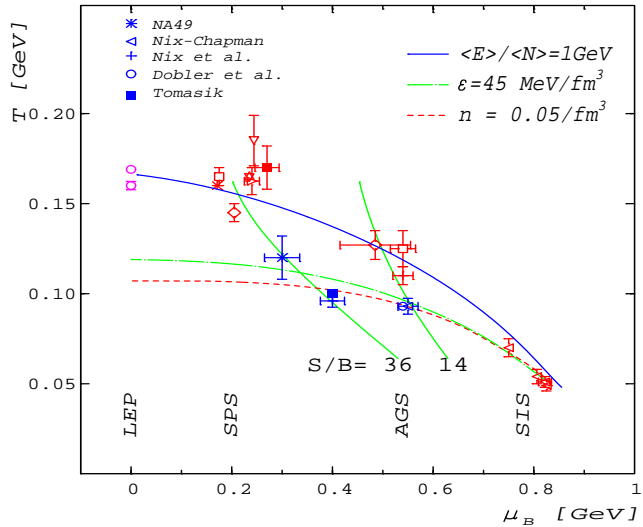}}
\caption{Chemical and thermal freeze-out points \protect \cite{clered}
in the $(T, \mu_B)$ plane for various experiments.} 
\label{fig:two}
\end{figure}

Assuming that a thermal freeze-out is triggered by a rapid
expansion, one expects the momentum spectra of various hadrons to
reflect the freeze-out temperature, $T_{fo}$, which will be
blue-shifted by the collective expansion. For small transverse momenta, 
$p_T \ll m$, we expect the inverse slope of transverse mass
distribution, $d\sigma / d (m_T - m)$ with $m^2_T = p^2_T + m^2$,
to be given by $T_{\rm slope} = T_{fo} + {1\over 2} m \langle v_T
\rangle^2$. Thus one expects, $T_{\rm slope}$ to vary linearly with the
mass of the observed particle, $m$.  Such a linear rise has been seen in 
the Pb-Pb data, leading to an estimate for $T_{fo}$. On the other hand,
the same transverse momentum distributions can also be equally well 
explained \cite{nick} using a non-thermal model, indicating
the non-uniqueness of the thermal interpretation and hence of $T_{fo}$.

Strangeness changing -- chemical-- reactions are typically slower than
the elastic processes and hence are expected to freeze-out before the
thermal freeze-out. The temperature and chemical potential at this
freeze-out decides the particle yields of various types, provided these
yields are measured for the full 4$\pi$-integrated region; otherwise the
measurements will depend upon the details of the collective flow
mentioned above. Furthermore, taking ratios of such yields, one can
reduce the dependence on the collective dynamics even more. A simple
thermal model of free particles at a temperature $T$, volume $V$ and
chemical potential $\mu_B$ has been shown \cite{braun} to describe
beautifully 22 ratios of particle yields which vary by three
orders of magnitude, leading to $T^{chem}_{fo} \simeq 170$ MeV and
$\mu_{B,fo}^{chem} \simeq 270$ MeV. Fig. \ref{fig:two} displays the thermal
and chemical freeze-out points for the SPS $Pb$-$Pb$ collisions along
with those of other experiments.  A comment about
$\mu^{thermal}_{B,fo}$ may be in order, since I discussed above only the
corresponding $T_{fo}$. As chemical equilibrium is lost earlier,
it is strictly speaking not well defined. One simply adjusts 
$\mu_{B,fo}^{thermal}$ such that the particle ratios at 
$T^{therm}_{fo}$ agree with the observed values.

Since $T^{chem}_{fo}$ turns out to be very close to that expected for
the quark-hadron transition from lattice QCD, it is plausible that the
hadronic chemical equilibrium is a direct consequence of a pre-existing state 
of uncorrelated quarks and antiquarks and not due to hadronic
rescatterings/reactions, since there is not much time for the
latter. Hadron formation is then governed by the
composition of the earlier state in a statistical manner and an expansion 
later does not change their yields. Needless to say though, the
proximity of the two temperatures mentioned above is only
suggestive. Indeed such temperatures and chemical potentials could
still be reached via an expanding hadron gas as well.   One then would
expect though that the particle yield ratios will not reflect the underlying
quark symmetries, as have been seen in the strangeness
enhancement \cite{wa97} pattern observed by the WA97 collaboration.

\subsection{$J/\psi$ Suppression}

As remarked in the introduction above, one needs to employ harder
probes to explore the physics of the fireball at earlier times when
QGP may have existed. Production of $J/\psi$ is one such hard
probe. Since it is a tightly bound meson of charm and anticharm quarks, Matsui 
and Satz \cite{helmut} argued that color Debye screening of these heavy
quarks will prevent formation of $J/\psi$, if QGP is formed in the
heavy ion collisions. Due to a finite size and lifetime of the fireball, the
observable effect is expected to be a suppression in the production of
$J/\psi$. The NA38 and NA50 collaborations \cite{na50} measured
$J/\psi$ cross sections for a variety of collisions, starting from
$p+d$ to $Pb + Pb$ using the same muon spectrometer in the same
kinematic domain ($0 \leq y^{cm}_{\mu^+\mu^-} \leq 1$ and
$|\cos\Theta_{cs}| \leq 0.5$). While the systematic errors are thus
minimised, the lighter beams were necessarily of high energies;
$\sqrt{S_{NN}}$ thus varies from 17 GeV to 30 GeV. 

\begin{figure}[htbp]
\epsfxsize=9cm
\epsfysize=9cm
\centerline{\epsfbox{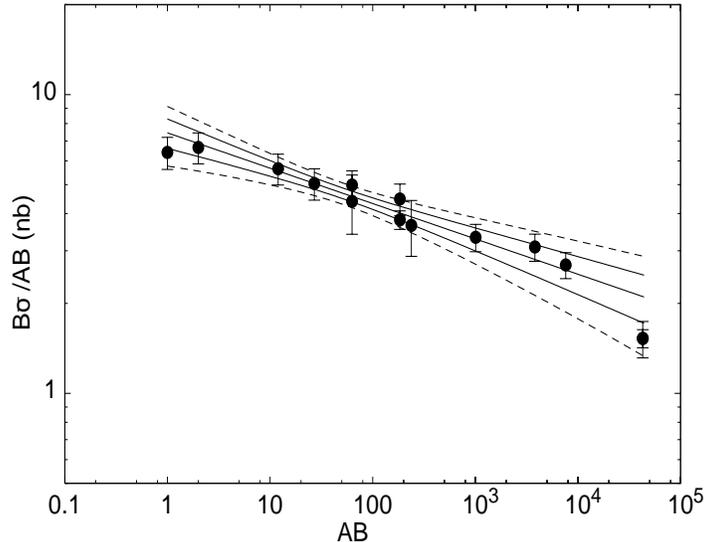}}
\caption{ The data \protect \cite{na50} for $J/\psi$ cross section times 
its branching fraction into dimuons as a function of product of the mass 
numbers of target and projectile AB with the 1$\sigma$ (full lines) and
2$\sigma$(dashed lines) theoretical predictions\protect  \cite{us2}.} 
\label{fig:four}
\end{figure}

Comparing the $\sigma^{DY}_{obs}$ with $\sigma^{DY}_{LO,th}$ a universal 
$K$-factor was found in $pp$, $pA$ and $AB$ collisions: $\sigma^{DY}_{A\cdot
B} \propto A \cdot B$ for all of them, where $A$ and $B$ are the mass
numbers of the projectile and target respectively. Normalizing
$B_{\mu^+\mu^-} \sigma^{J/\psi}_{AB}$ by dividing by $A \cdot B$
therefore, where $B_{\mu^+\mu^-}$ is the branching fraction of
$J/\psi$ in to $\mu^+\mu^-$, one could expect QGP formation to be
signalled by a drop at some value of $A \cdot B$. Fig. \ref{fig:four}
shows the NA38 and NA50 results where one notices a gradual fall in with $A
\cdot B$ for {\it all} values. Note that some measurements have been re-scaled 
so that all are for the same energy in this figure. The decreasing cross
section for all values of $A \cdot B$, including small ones, is an indication 
of the presence of yet another mechanism for $J/\psi$-suppression in these 
collisions. Thus any suppression due to QGP will have to be over and above 
this `normal suppression'.

Production of heavy quarkonia is an old and mature area of
perturbative QCD. In particular, hadroproduction of $J/\psi$ has been
explained both in the colour evaporation model \cite{us1} and the
colour octet model \cite{gupsri} at $\sqrt{s}$ comparable to those in
Fig. \ref{fig:four}. So it is a natural question to ask whether 
the decrease in Fig. \ref{fig:four}
can be explained using pQCD. Unfortunately, sufficient information on
the gluonic nuclear structure functions is not
available at present; assuming them to be independent of mass number
$A$ or $B$ is perhaps incorrect in view of the famous EMC-effect. 
Using the existing models of the EMC-effect, on the other hand, one finds 
hardly any decrease in the cross section in Fig. \ref{fig:four}. 
The lack of decrease of $B_{\mu^+\mu^-} \sigma^{J/\psi}_{AB} /AB$ with $AB$ 
appears to be a generic feature, since the dominant 
contribution to the cross section in Fig. \ref{fig:four} comes from the
so-called anti-shadowing region in $x$ which ought to be there for even the
gluons due to the momentum sum rule.  In view of the continuous decrease in
Fig. \ref{fig:four}, i.e. even for $p +$ light-$A$, where the radius of 
the target is only 2-3 times larger than that of proton, i.e, for 
hadroproduction \cite{us1,gupsri}, one has to ask whether a 
pQCD description of total cross sections for $J/\psi$ is at all
possible. It would be interesting and desirable to thrash out this 
question by extensive investigation of the nuclear glue and its impact 
on the $J/\psi$ cross section.

The normal suppression in Fig. \ref{fig:four} has been 
explained \cite{abs} as a final state interaction. The produced $J/\psi$-state 
or its precursor can get absorbed in the nuclear matter (of the target and
beam). Treating $\sigma^{\psi N}_{abs}$ as a free parameter and using
the known nuclear profiles, one finds that a $\sigma_{abs} \sim$ 6.4 mb
can explain the linear fall in Fig. \ref{fig:four} quantitatively in
Glauber type models.  However, the $Pb$-$Pb$ data point seems to be off
this linear fall, and exhibits thus an `anomalous suppression'. One can
alternatively use an empirical $(AB)^\alpha$ fit to all points except
the $Pb$-$Pb$, which too will be linear on the scales of Fig.
\ref{fig:four}, and the $Pb$-$Pb$ data point stands out again.

Unfortunately, the issue of how statistically significant this anomalous
suppression is gets affected by the crudeness of the theory described
above as well as by the assumptions needed to rescale some of the data
points. Ignoring these systematical theoretical errors, one finds the
anomalous suppression to be a 5$\sigma$ effect \cite{na50}, while
including them leads \cite{us2} to a conclusion that no anomalous
suppression exist at a 2$\sigma$ or 95\% confidence level, as shown by
the 2$\sigma$-band (enclosed by dashed lines) in Fig. \ref{fig:four}.

The NA50 collaboration also measures $J/\psi$-suppression as a function of 
the total produced transverse energy $E_T$.  By taking the ratio of the 
number of $J/\psi$ events and the Drell-Yan events in each $E_T$-bin, 
one obtains a less systematic error prone $R_{\rm expt} = B_{\mu^+\mu^-}
\sigma^{J/\psi}/\sigma^{DY}_{M_1-M_2}$ as a function of $E_T$, where
$M_1$-$M_2$ is the range of dimuon mass over which the Drell-Yan cross
section is integrated.  Using simple geometrical models, $E_T$ can be 
related to the impact parameter $b$ at which the two nuclei collide. 
Furthermore, any given $b(E_T)$ can be related to an average nuclear
path length $L$ which the produced $J/\psi$ (or its precursor) has to
traverse and which will determine the probability of its absorption in
nuclear matter.

\begin{figure}[htbp]
\epsfxsize=9cm
\epsfysize=8cm
\centerline{\epsfbox{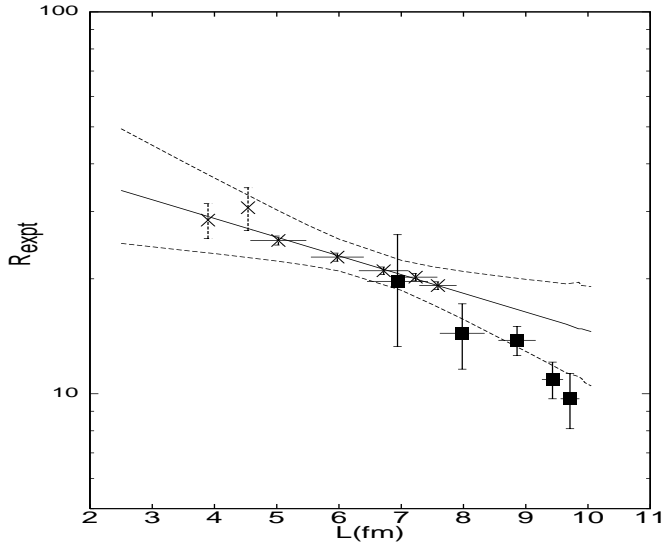}}
\caption{ The data \protect  \cite{na50} from NA38 (crosses) and NA50
(squares) with 4$\sigma$ errors for the ratio of $J/\psi$ cross section 
and the Drell-Yan cross section as a function of $L$ in fm along with a 
theoretical fit and a 4$\sigma$ band \protect  \cite{us3} around it.} 
\label{fig:five}
\end{figure}

Fig. \ref{fig:five} shows $R_{\rm expt}$ as a function of $L$, as
determined by the NA50 collaboration, using $M_1$ = 2.9 and $M_2$ = 4.5. 
The normal nuclear suppression can be well approximated by $R_{\rm expt}
= A \cdot \exp (- \rho_{nucl} \cdot \sigma_{abs} \cdot L)$ or can be 
calculated more exactly in a Glauber model. The straight line in Fig.
\ref{fig:five} displays the fit for the light nuclei for $\rho_{nucl} =
0.17/$ fm$^3$ and $\sigma_{abs} \simeq 6.6$ mb. The low $L$ point
for $Pb$-$Pb$ collisions, corresponding to peripheral collisions, falls
on the fitted line while all the large $L$ points fall below it. Again,
one can ask for the statistical significance of this anomalous
behaviour. Since the fit above uses data from $E_T$-bins, or
equivalently $L$-bins, for lighter nuclei, there are again sizeable
errors on the theoretical prediction. For the 1995 data, which seem
broadly in agreement with the 1996 data and the latest 1998 data, it has been
estimated \cite{us3} that all the $Pb$-$Pb$ data points fall in a
4$\sigma$-band although they are all systematically below the
theoretical prediction, as shown in Fig. \ref{fig:five}.

It seems thus likely that an additional mechanism to suppress $J/\psi$ 
production in $Pb$-$Pb$ collisions is needed over and above the normal
nuclear absorption.  This is even more so for the second shoulder in 
the $E_T$-spectrum, observed \cite{NewNa50} in the 1998 data.  There 
have been several theoretical attempts to provide such a mechanism 
including a possible a quark-hadron transition.  A key non-QGP 
scenario invokes the possibility of destruction of the $J/\psi$ by the
so-called co-mover debris of the collisions.  While the second shoulder
was anticipated \cite{GuSa} in a QGP model, it has been explained \cite{cap}
in the co-mover picture as due to fluctuations in the tail of the 
$E_T$-spectrum.  The difference between the two models 
may, therefore, show up at the upcoming RHIC collider in BNL where 
$Au$ (19.7 TeV) + $Au$ (19.7 TeV) collisions will be studied this year
and the $E_T$-tail will extend much farther, although in another
QGP-like model \cite{blaizot} fluctuations in $E_T$ have been argued
to explain the second shoulder in the NA50 data successfully.  

\section{Conclusions and Outlook}

An important non-perturbative prediction of (lattice) QCD is the
existence of a new phase of matter, Quark-Gluon Plasma, at
sufficiently high temperatures. Since the Standard Model has so far
been tested experimentally only in the weak coupling regime, it seems
desirable to confront this prediction with experiments. Collisions of
heavy ions at very high energy may be able to deposit the
required high energy density over a reasonable volume. The
experimental programs at BNL, New York and CERN, Geneva have by now
provided results for $Au$ on $Au$ and $Pb$ on $Pb$ at $\sqrt{s}
\simeq$ 0.9 TeV and 3.6 TeV (or $\sqrt{s}_{NN} \simeq$ 5 GeV and 17
GeV) respectively. The year 2000 should witness a factor of about 39
increase in the colliding CMS energy at BNL while LHC at 
CERN should achieve a $\sqrt{s}$ = 1150 TeV. The experiments so far have 
provided tantalizing hints of the new phase and therefore of the exciting
physics in the years ahead.

A fireball of QGP produced in these collisions cools by expanding
and converts into ordinary hadrons and leptons fairly quickly. Since
this makes a distinction of events with QGP formation from those without
it a very tough task, it seems prudent to look for a congruence of various
signatures in as many different ways of detecting QGP as
possible. The current results do indicate such a trend.
Soft hadron production data can be interpreted
in terms of a chemical freeze-out followed by a thermal
freeze-out. The freeze-out temperature for the former for the CERN SPS 
data turns out to be $\sim$ 170 MeV $\simeq T_c$ (quark-hadron
transition).  The strangeness enhancement pattern 
seen by the WA97 experiment, showing larger
enhancement for the heavier particles with more strange quarks, 
also suggests that the hadrons at chemical freeze-out were formed from an
uncorrelated QGP-like state. 

Finally, anomalous $J/\psi$ suppression seen by the NA50 experiment
for $Pb$-$Pb$ collisions can be understood as arising out of a
deconfined quark-gluon plasma. Nevertheless, much more theoretical and 
experimental work will be needed to make a convincing case of
quark-gluon plasma formation in the heavy ion experiments since the
signals are still not spectacular in their statistical significance and
credible alternative explanations exist in many cases for the observed
results . Clearly, the commissioning of RHIC will be a big boost and will 
hopefully result in making a definitive case for quark-gluon plasma.

\end{document}